\documentclass[intlimits,twoside,a4paper]{article}

\usepackage{graphicx}
\usepackage[T2A]{fontenc}
\usepackage[cp1251]{inputenc}

\usepackage{cmpj2}

\issue{2013}{16}{4}{43607}
\doinumber{10.5488/CMP.16.43607}

\title[Density anomaly. Monte Carlo simulations]
{Density anomaly of charged hard spheres of different diameters
in a mixture with core-softened model solvent. Monte Carlo
simulation results}

\author[B. Hribar-Lee, O. Pizio]{B. Hribar-Lee\refaddr{label1},
          O. Pizio\refaddr{label2}\thanks{E-mail: pizio@unam.mx}}
\addresses{
\addr{label1}Faculty of Chemistry and Chemical Technology,
University of Ljubljana, \\ A\v{s}ker\v{c}eva c.\ 5, SI-1000 Ljubljana, Slovenia,
\addr{label2}Instituto de Qu\'{i}mica, Universidad Nacional Aut\'{o}noma de M\'{e}xico,
Circuito Exterior, 04510, M\'{e}xico, D.F., M\'{e}xico}

\date{Received August 7, 2013, in final form September 12, 2013}
\authorcopyright{B. Hribar-Lee, O. Pizio, 2013}

\begin{document}
\maketitle

\begin{abstract}
Very recently the effect of equisized charged hard sphere
solutes in a mixture with core-softened fluid model
on the structural and thermodynamic
anomalies of the system has been explored in detail
by using Monte Carlo simulations and integral equations theory
[J. Chem. Phys., 2012, \textbf{137}, 244502].
Our objective of the present short work is to complement this study by
considering univalent ions of unequal diameters in a mixture with
the same soft-core fluid model. Specifically, we are interested in the analysis
of changes of the temperature of maximum density (TMD) lines with ion
concentration for three model salt solutes, namely
sodium chloride, potassium chloride and rubidium chloride
models. We resort to Monte Carlo simulations for
this purpose.  Our discussion also involves the dependences of
the pair contribution to excess entropy and of constant volume heat
capacity on the temperature of maximum density line. Some
examples of the microscopic structure of mixtures in question
in terms of pair distribution functions are given in addition.


\pacs 61.20.-p, 61.20.Gy, 6120.Ja, 61.20.Ne, 61.20.Qg, 65.20.-w

\end{abstract}

\section{Introduction}

It is our honor and pleasure to make contribution to this special issue
dedicated to brilliant scientist Prof. Myroslav Holovko on the occasion
of his 70th birthday.
During last years we were influenced by his developments and profound results
along different lines of research in the theory of liquids and mixtures,
interfacial phenomena and phase transitions.
We appreciate his mentorship in some projects, many useful pieces of advice and
friendly discussions concerning topics of our common interest.
During many years Myroslav Holovko explored electrolyte solutions
in very detail, see e.g., the classical monograph~\cite{myroslav1} written by him
together with his teacher Prof. I. Yukhnovskyi that summarizes their
effort prior to 1980. Later results can be found in many articles and reviews.

The principal objective of the present work is to complete one
extension of our very recent study of solutions comprising solute ions and
a soft-core model fluid as a solvent~\cite{miha1}.
During last decade  many studies in the theory of fluids have been
focused on single-component, isotropic core-softened  models
in which the repulsive part of the interparticle
interaction potential is peculiar (``softer'') at
``intermediate'' interparticle separations while a hard-core repulsion
acts at short distances.
This region of peculiar dependence of the pair interparticle interaction
is often  described  by linear or nonlinear
ramp~\cite{hemmer,jagla1,jagla2,jagla3,jagla4,jagla5}, by a
shoulder~\cite{franzese1,franzese2} or Gaussian term~\cite{barros1,barros2,barros3},
or another more sophisticated functional dependence, see e.g., recent review~\cite{chaim} and
references therein. A more complete list of references has been given in~\cite{miha1}
as well.

Unfortunately, the complex shape of the
interaction in this type of models requires application of solely numerical procedures for
their solution. However, peculiar structural, thermodynamic and dynamic properties
relevant to certain real liquids can be reached.
Models of single-component fluids with isotropic core-softened
potential have been studied by computer simulations in numerous works,
see the list of references in~\cite{miha1}.
For purposes of the present work it is necessary to mention that
the fluids with spherically symmetric core-softened potentials
exhibit anomalies in thermodynamic and dynamic
properties in close similarity to those observed in liquids with
directional inter-particle interactions, such as water, silica,
phosphorous, and some liquid metals.
One of the examples of such anomalous behavior is the presence of the fluid density
maximum as a function of temperature.
Moreover, some of the soft-core fluid
models exhibit puzzling liquid-liquid coexistence envelope
terminating at the critical point, in addition to common liquid-vapor
phase transition. Thermodynamic and other anomalies can be related to
this additional phase transition, see e.g., references~\cite{evy1,evy2,evy3}.
The soft-core models by design miss the
orientational interparticle correlations.
Thus, their correspondence
to, for example,  water or aqueous solutions is restricted to
thermodynamic properties.
Nevertheless, the soft-core models are simple and permit
to capture the peculiarities of thermodynamic properties quite
adequately~\cite{chaim}.

Theoretical studies of solutions in which a model solvent exhibits
thermodynamic properties with anomalies have been attempted
recently by different authors,
see e.g.,~\cite{pablo1,pablo2,buldyrev1,buldyrev2,buldyrev3,buldyrev4,egor3}.
Several important particular issues were explored, namely the
solubility of apolar solutes in such solvents, the effect of
solute species on the
temperature of maximum density line, on heat capacity extrema and others. On the other hand, following our previous
theoretical and simulation developments~\cite{pizio1,miha2},
we applied similar modelling, but involved
ionic solutes rather than nonpolar species~\cite{miha1}.
We examined the effect of
ionic solutes on the microscopic structure and thermodynamics of
solutions with a soft-core solvent that possesses anomalies in one-phase
region.
Experiments~\cite{archer1,archer2,mishima1,mishima2} and computer
simulations~\cite{gallo1,gallo2,ludwig}
have shown that certain anomalies of thermodynamic properties
are preserved in ionic aqueous solutions for low and up to
moderate ion concentrations. They become weaker and finally
disappear, if the concentration of ions increases but remains below
the solubility limit.

In contrast to the initial stage of this project~\cite{miha1},
we consider a mixture of singly charged cations and anions
modelled as charged hard spheres and solvent particles
described by a soft-core model, all of them having
nonequal diameters. The system is studied by the canonical Monte Carlo simulations.
The following discussion of the model and methods is brief in order to avoid
unnecessary repetitions of~\cite{miha1}, only the essential elements are given
to benefit the reader.

\section{Model and method}

The system in question consists of ionic solute (salt) and solvent. The
ionic subsystem is a mixture of positively and
negatively charged hard spheres with arbitrary diameters, $\sigma_+$ and $\sigma_-$, the charge is
located at the particle center. Ions and solvent particles are immersed in a
dielectric background characterized by dielectric constant
$\varepsilon$. The interaction potential between ions $i$ and $j$ is,
\begin{eqnarray}
U_{ij}(r)=\left\{
\begin{array}{ll}
\infty \,, & \hbox{$r < (\sigma _{i} + \sigma _{j})/2$},\\
\displaystyle \frac{e^2z_iz_j}{4\pi\varepsilon_0 \varepsilon r}\,, &
\hbox{$r \geqslant (\sigma _{i} + \sigma_{j})/2$},
\end{array}
\right.
\label{eq:coulomb}
\end{eqnarray}
where $e$ is the elementary charge, $z_i = |z_-| = 1$ are the valencies of ions,
$\varepsilon_0$ is the permittivity of free space, and $r$ is the inter-particle distance.
As in~\cite{miha1}, we use the core-softened model by Barros de Oliveira et al.
\cite{barros1,barros2,barros3} that possesses thermodynamic, structural, and dynamic
anomalies. The potential is a sum of the Lennard-Jones and a Gaussian terms,

\begin{equation}
U_{\mathrm{ss}}(r)= 4\varepsilon_\mathrm{ss} \left [ \left (\frac{\sigma_\mathrm{s}}{r} \right )^{12}
- \left (\frac{\sigma_\mathrm{s}}{r} \right )^{6} \right ] +
a \varepsilon_\mathrm{s} \exp \left [ - \frac{1}{c^2}\left (\frac{r-r_0}{\sigma_\mathrm{s}} \right )^2 \right ],
\label{eq:solvent}
\end{equation}
with $\varepsilon_\mathrm{ss}$~--- the solvent-solvent interaction energy
and $\sigma_\mathrm{s}$~--- the diameter of a solvent particle.
Parameters $a$, $c$, and $r_0$ are taken as in the original work~\cite{barros1}
($a = 5$, $c=1$, and $r_0/\sigma_\mathrm{s} = 0.7$). With this choice of parameters,
the inter-particle attraction is strongly depressed and thus there is no liquid-gas
transition in the solvent subsystem~\cite{dasilva}.

The pressure-temperature curves calculated for the pure solvent
at certain densities have a minimum (i.e., $(\partial P/\partial T)_{\rho}=0$)
and consequently $(\partial \rho/\partial T)_P$ vanishes for some thermodynamic
states. The region of density anomaly corresponds to the state
points for which $(\partial \rho/\partial T)_P > 0$ and is
bounded by the locus of points for which the thermal expansion
coefficient equals zero.

There is some room for the choice of ion-solvent interaction. However, to be
consistent with our previous work~\cite{miha1},
the interaction between an ion ($i$), and a solvent ($s$)
particle is taken simply in the form of Lennard-Jones (12-6) potential,
\begin{equation}
U_\mathrm{is}(r)=4\varepsilon_\mathrm{is} \left [ \left (\frac{\sigma_\mathrm{is}}{r} \right )^{12}
- \left (\frac{\sigma_\mathrm{is}}{r} \right )^{6} \right ],
\end{equation}
where $\varepsilon_\mathrm{is}$ is the ion-solvent interaction energy
and $\sigma_\mathrm{is}= (\sigma_\mathrm{i} + \sigma_\mathrm{s})/2$.  The reduced temperature for the
solvent subsystem is defined as $T^* = k_\mathrm{B} T / \varepsilon_\mathrm{s}$,
where $k_\mathrm{B}$ is the Boltzmann constant and $T$ is the absolute temperature.

In order to describe the energetic aspects of interactions between all species,
it is necessary to
define the parameter $\alpha = T^*/T^*_\mathrm{el}$ as the ratio
between the reduced temperature for the solvent subsystem ($T^*$), and the reduced temperature of the ionic subsystem,
$T^*_\mathrm{el} = \sigma_{-}/\lambda_\mathrm{B}$, where
$\lambda_\mathrm{B} = e^2/(4\pi \varepsilon_0 \varepsilon k_\mathrm{B} T)$ is the
Bjerrum length, $\lambda_\mathrm{B} = 7.14$~\AA.
Similar to the previous study~\cite{miha1}, we choose $\alpha =
0.25$. Thus, if the solvent is even at a rather low temperature,
the ionic system remains far from the possible vapor~-- liquid type criticality.
The mixing rule for ion-solvent potential is applied such that:
$\varepsilon_\mathrm{is}= (\varepsilon_{ii}\varepsilon_{jj})^{1/2}$,
where $\varepsilon_\mathrm{ss}=1/T^*$, $\varepsilon_{--}=1/T^*_\mathrm{el}$ and
$\varepsilon_{++}=(\sigma_-/\sigma_+)/T^*_\mathrm{el}$.

We performed Monte Carlo computer simulations in the canonical
ensemble and in the cubic simulation box according to common
algorithms~\cite{allen,frenkel}.
The simulation was started from a configuration of randomly
distributed particles.
The number of ions in the box, $N_\mathrm{ion}$ ($N_\mathrm{ion}=N_+ + N_-$)
was chosen to provide their reasonable statistics and along with the desired
ion concentration, $x_\mathrm{ion}$ [$x_\mathrm{ion}=N_\mathrm{ion}/(N_\mathrm{ion}+N_\mathrm{s})$], where
$N_\mathrm{s}$ is the number of solvent particles resulting from $N_\mathrm{ion}$
and $x_\mathrm{ion}$.
In addition, we picked up the desired
total dimensionless density of the system, $\rho^*$,
$\rho^*=(N_+\sigma_+^3 + N_-\sigma_-^3 + N_\mathrm{s}\sigma_s^3)/L^3$,
where $L$ is the length of the box edge resulting from previously
given parameters. Coulomb
interactions were taken into account by using the Ewald
summation ~\cite{allen,frenkel} with 337 wave vectors and
the inverse length equal to $5/L$).

In this work, we explore three model salts in the soft-core
solvent, namely the NaCl ($\sigma_+=2.21$~\AA, $\sigma_-=4.42$~\AA),
KCl ($\sigma_+=4.74$~\AA, $\sigma_-=4.42$~\AA) and
RbCl ($\sigma_+=5.27$~\AA, $\sigma_-=4.42$~\AA)
see e.g.,~\cite{balbuena,mountain}.
In all cases, we assume $\sigma_\mathrm{s}= 3.166$~{\AA} in close similarity to the SPCE model of water.
Each of the systems was first equilibrated over $10^8$
attempted configurations followed by a series of production runs of at
least $5 \times 10^7$ attempted particle displacements to
obtain the relevant average properties of the system under
investigation (excess internal energy, compressibility factor,
heat capacity at constant volume, pair excess entropy, and
radial distribution functions). The number of ions present in
the simulation box was in the range between $150$ and $400$ (dependent on the
desired density). The excess internal energy of the
system was obtained directly as an ensemble average. The
equation of state was calculated by~\cite{hansen}
\begin{equation}
\frac{PV}{Nk_\mathrm{B}T} = 1 - \frac{1}{3N} \left \langle r \frac{\partial \beta U_\mathrm{tot}}{\partial r} \right \rangle
+ \frac{2\pi}{3} \sum_{i} \sum_{j}
 x_i\frac{N_j}{V} \sigma_{ij}^3 g_{ij}(\sigma_{ij}) \,, \qquad (i,j=+,-),
\label{eq:osmotic}
\end{equation}
where $U_\mathrm{tot}$ is the total excess internal energy of the
system and $g_{ij}(\sigma_{ij})$ is the contact value of the
appropriate ion-ion radial distribution function.
Pair excess entropy was determined from
the expression~\cite{truskett}
\begin{equation}
\frac{S_{2,\mathrm{ex}}}{k_\mathrm{B}} = -2 \pi \rho \sum_{i,i} x_i x_j \int \left [ g_{ij}(r) \ln g_{ij}(r) -
g_{ij}(r) + 1 \right ] r^2 {\rd}r \,,
\label{eq:s2}
\end{equation}
where $x_i$ is the number fraction of species $i$,
and $g_{ij}(r)$ are the pair distribution
functions of species $i$ and $j$. The heat capacity at constant volume was determined
from the energy fluctuations as
\begin{equation}
 C_V = \left(\langle U_\mathrm{tot}^2 \rangle - \langle U_\mathrm{tot} \rangle^2\right)/k_\mathrm{B} T^2 \,.
\label{eq:CV}
\end{equation}

\section{Results and discussion}

\begin{figure}[!b]
\centerline{
\includegraphics[width=0.415\textwidth,clip]{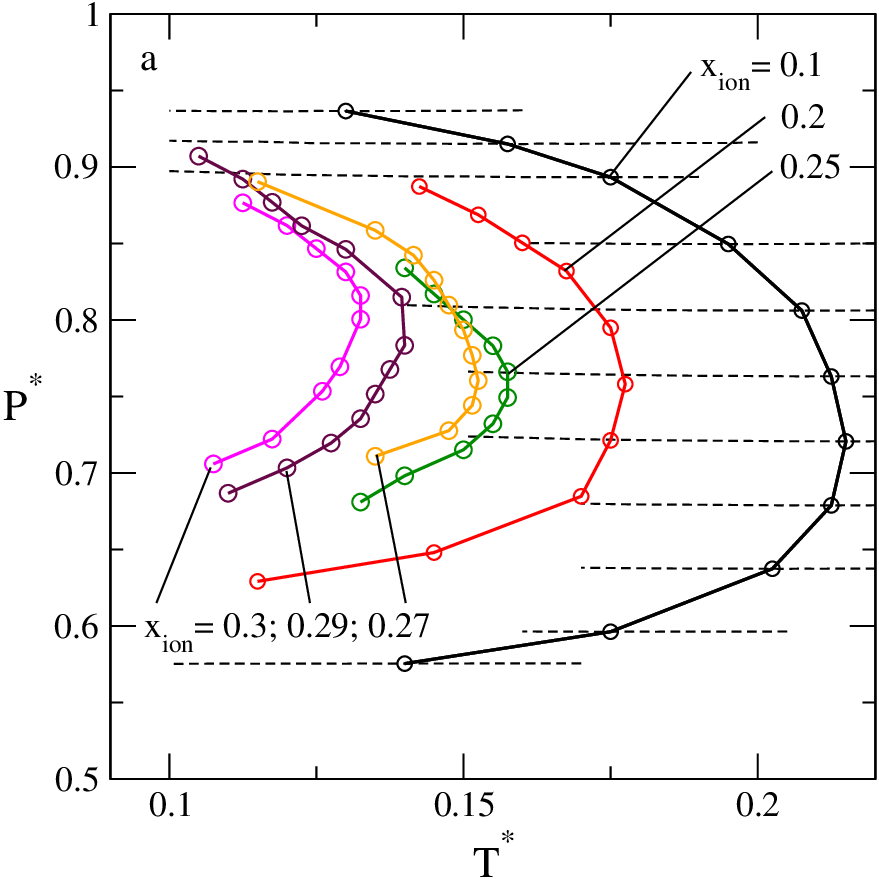}
\hspace{5mm}
\includegraphics[width=0.45\textwidth,clip]{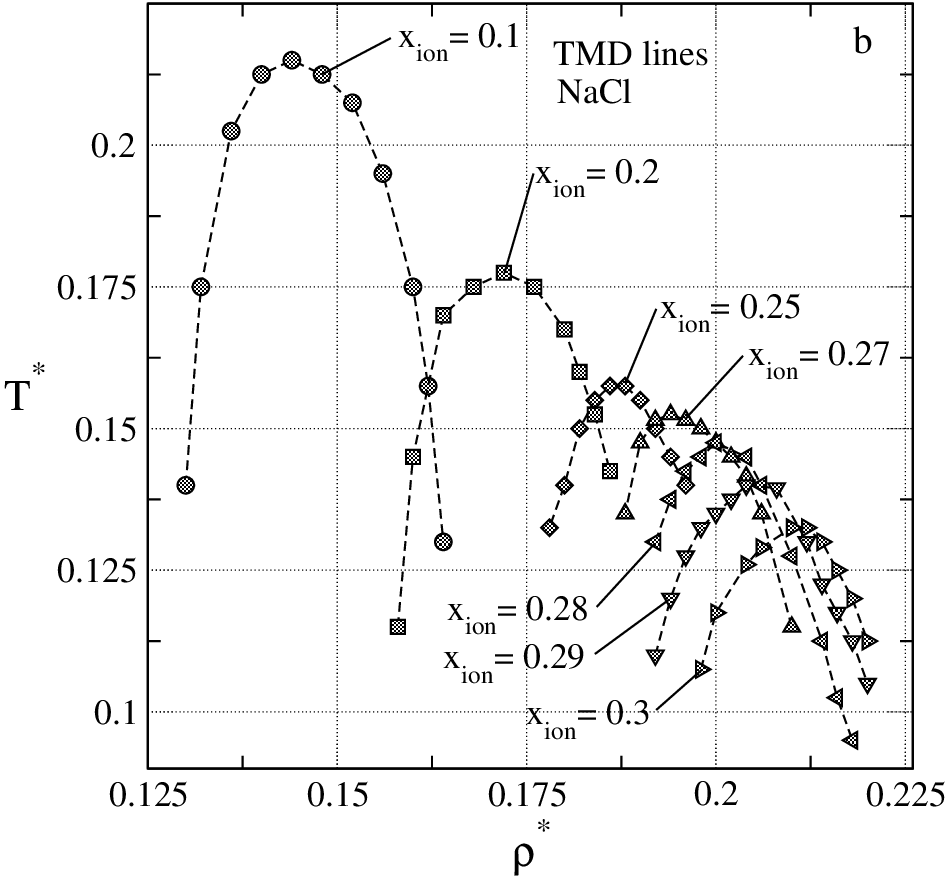}
}
\caption{(Color online) Temperature of maximum density lines of the {NaCl} model
in the $P^* - T^*$ plane (panel a)
$T^* - \rho^*$ plane (panel b).\label{f1}}
\end{figure}

\begin{figure}[!t]
\centerline{
\includegraphics[width=0.45\textwidth,clip]{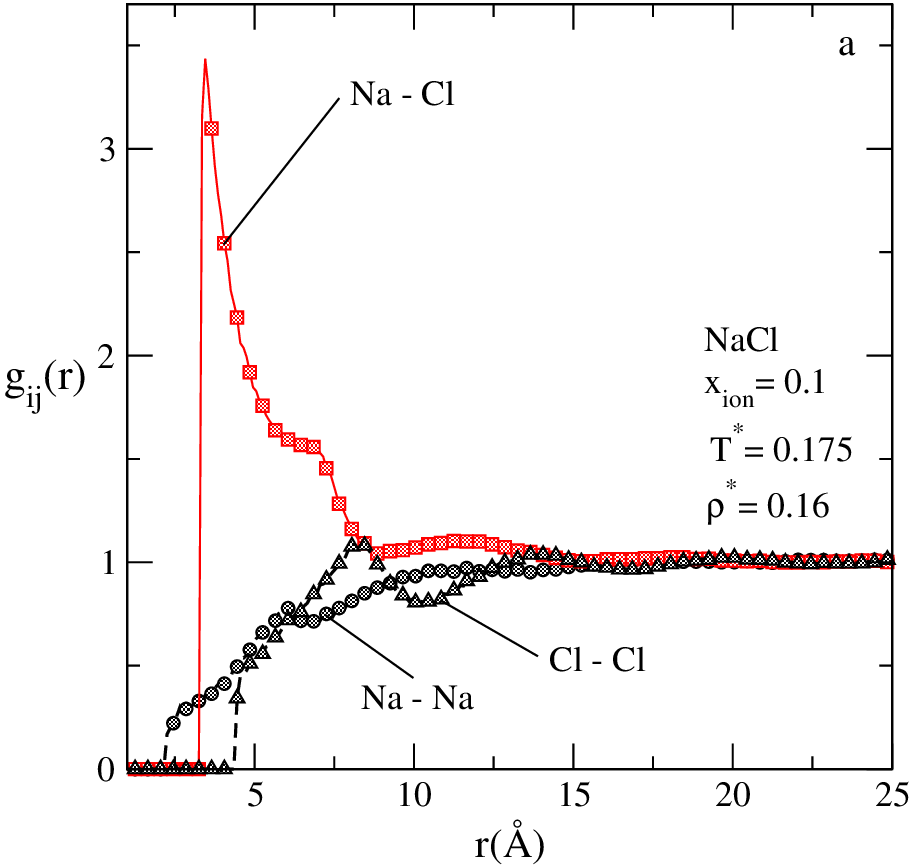}
\hspace{5mm}
\includegraphics[width=0.45\textwidth,clip]{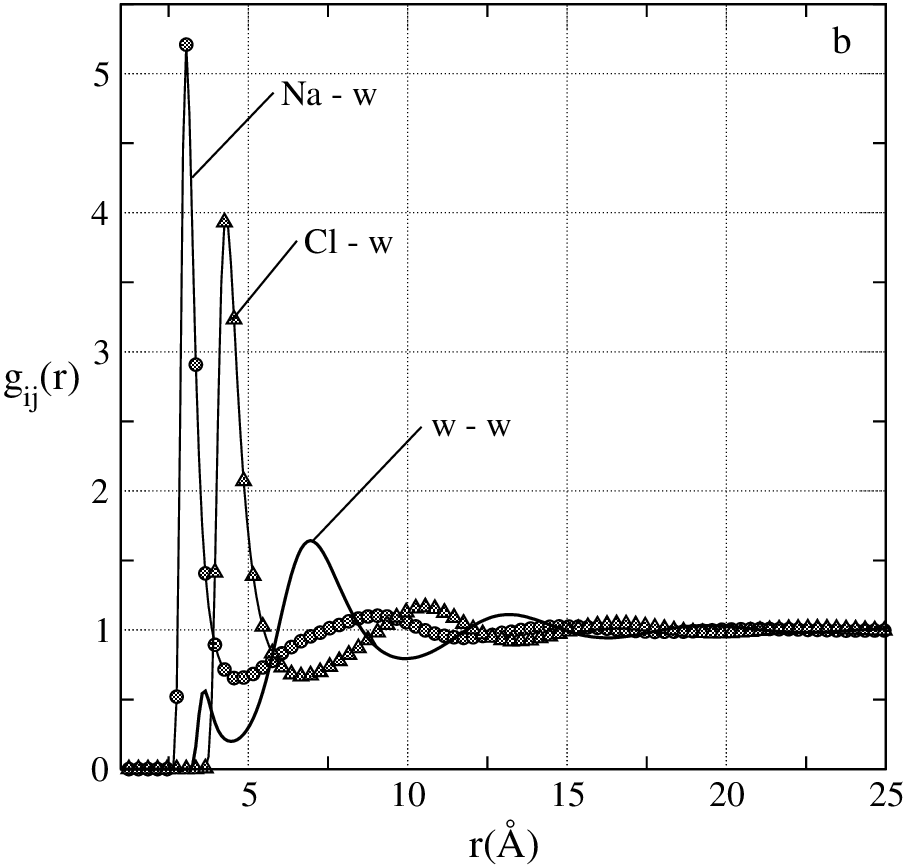}
}
\caption{(Color online) Pair distribution functions of the {NaCl} model. Parameters
of the system are given in the figure.\label{f2}}
\end{figure}

\begin{figure}[!b]
\centerline{
\includegraphics[width=0.45\textwidth,clip]{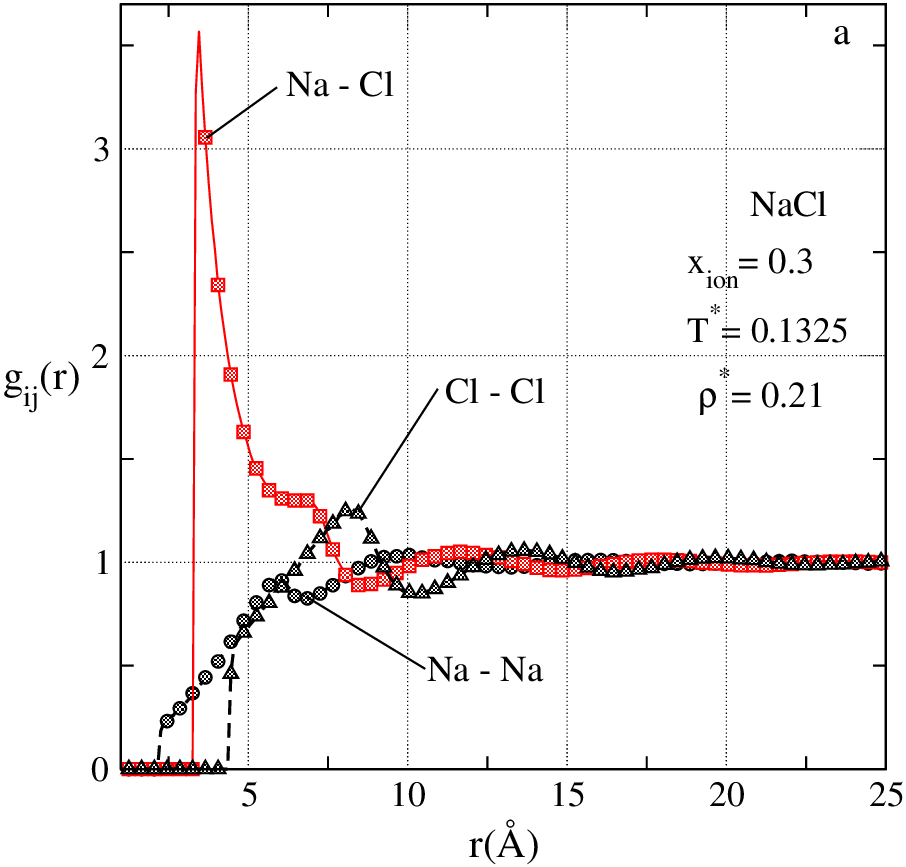}
\hspace{5mm}
\includegraphics[width=0.45\textwidth,clip]{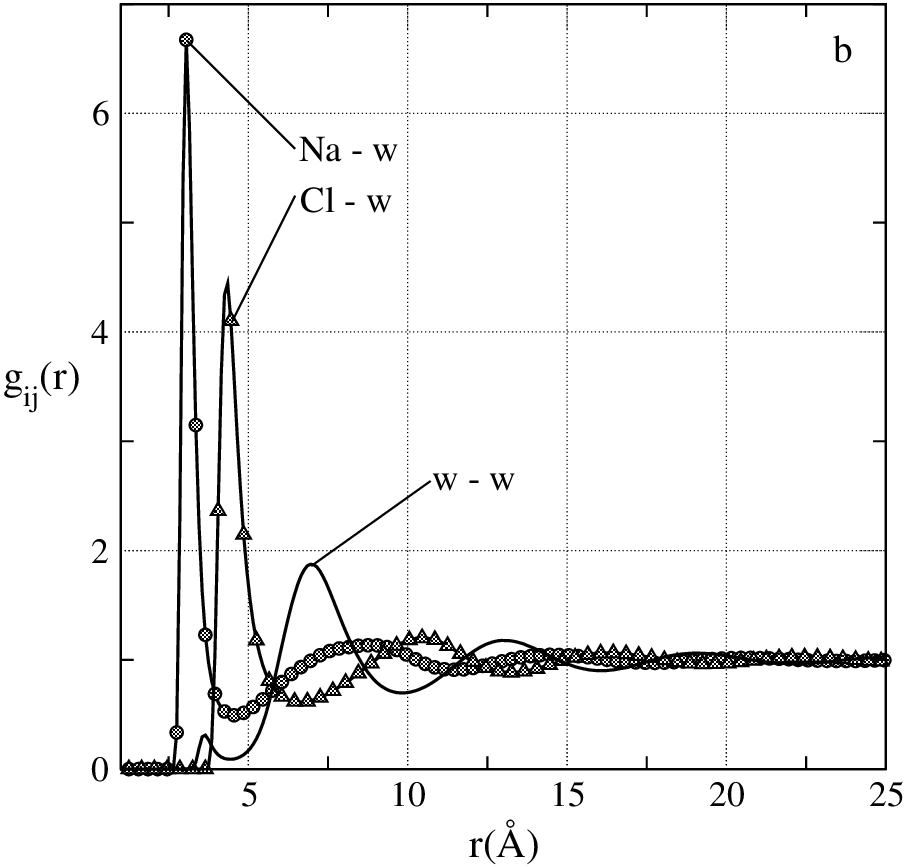}
}
\caption{(Color online) Pair distribution functions of the {NaCl} model. Parameters
of the system are given in the figure.\label{f3}}
\end{figure}

\begin{figure}[!t]
\centerline{
\includegraphics[width=0.45\textwidth,clip]{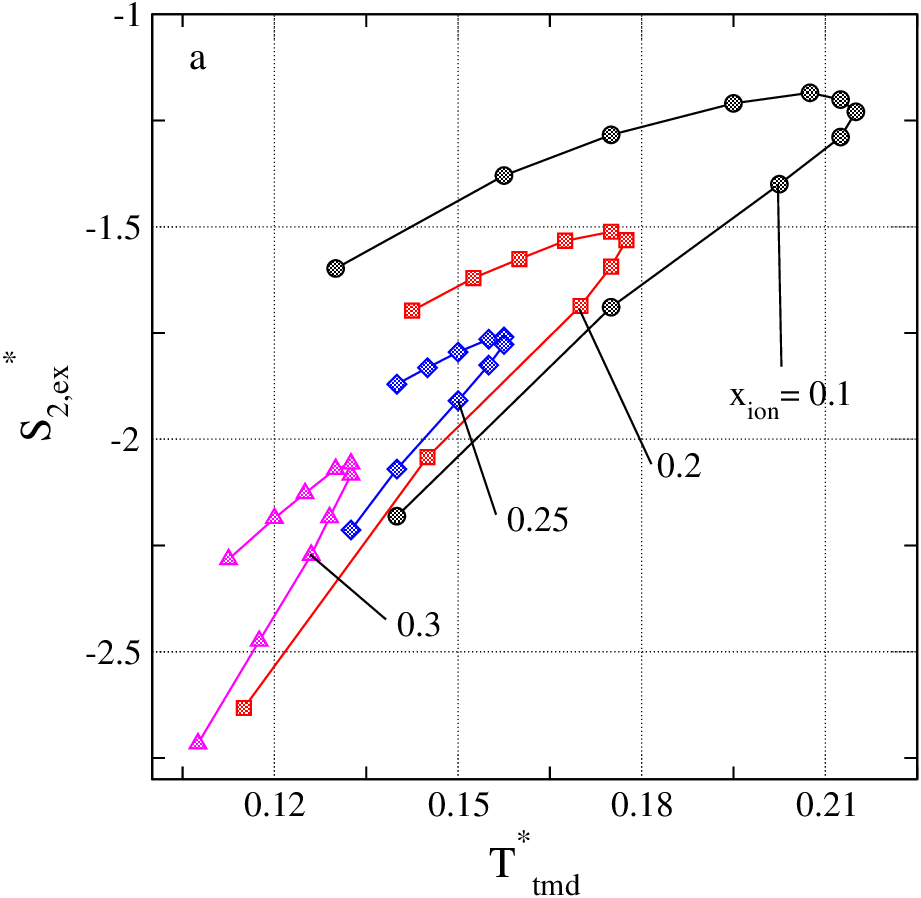}
\hspace{5mm}
\includegraphics[width=0.45\textwidth,clip]{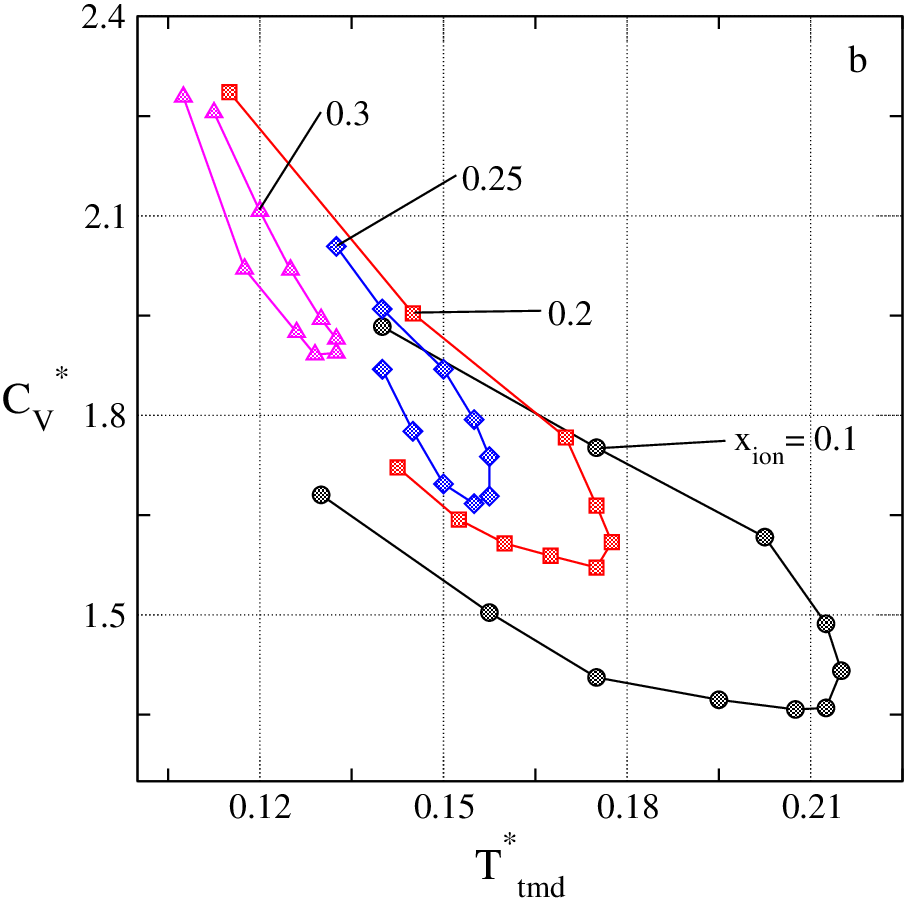}
}
\caption{(Color online) The behavior of the pair contribution to excess entropy (panel a)
and of reduced heat capacity (panel b) on the
temperature of maximum density of the {NaCl} model
at different concentration of ions.}\label{f4}
\end{figure}

First series of our calculations concern the NaCl salt model ($\sigma_+=2.21$~\AA, $\sigma_-=4.42$~\AA)
in the soft-core solvent. The cation is rather small and
its diameter is smaller compared to the anion and to the diameter of solvent
particles. The results for $P^*$ ($P^*=P\sigma^3_\mathrm{s}/\varepsilon_\mathrm{s}$)
as a function of $T^*$
for a set of densities of the mixture at fixed ion
concentration $x_\mathrm{ion}=0.1$ are given by short-dashed lines
in figure~\ref{f1}~(a). Each line has a minimum (it is difficult to see it
on the scale of such combined figure) marked by a circle. By
joining the circles, one can construct the temperature of maximum
density line at this particular ion concentration, $x_\mathrm{ion}=0.1$.
The $P^*(T^*)$ lines at a given $\rho^*$ are built by joining
points obtained as an average from several runs. Actually, it is
quite a tedious procedure. We perform a series of calculations to construct
the TMD lines for different $x_\mathrm{ion}$, up to $x_\mathrm{ion}=0.3$.
Each of the lines has an extremal point that shifts to a higher pressure
and lower temperature with an increasing $x_\mathrm{ion}$.
Moreover, the TMD lines shrink with an increasing
ion concentration. These trends were already discussed by us for the
model in  which all the species involved are of equal diameter~\cite{miha1}.
Slightly different view on the changes of the TMD lines is given in figure~\ref{f1}~(b).
The shift of the TMD lines to higher $\rho^*$ with a growing $x_\mathrm{ion}$
can be explained by augmenting the excluded volume effects in the system,
if ions are added to the solvent. On the other hand,
the extremum point of the TMD
along the temperature axis goes down due to the overall growth
of attraction, principally between oppositely charged ions.

Previously, we discussed the evolution of the microscopic
structure of the model characterized by equal diameters of all species
on the density at a fixed $T^*$ and $x_\mathrm{ion}$, see figures~1 and 2 of \cite{miha1}.
Trends of behavior of the pair
distribution functions are similar to the model with non-equal diameters.
Two examples of the distribution of particles at thermodynamic states
on the relevant TMD line are given here in figures~\ref{f2} and \ref{f3}.
In the case of a lower density and lower ion concentration (figure~\ref{f2}),
we can attribute the shape of the ion-ion distribution functions
to the formation of Na--S--Cl, Cl--S--Cl and Na--S--Na
structures or complexes, i.e., of ions separated by a single solvent
molecule, corresponding to the local maxima at $(\sigma_\mathrm{Na}+\sigma_\mathrm{Cl})/2+w_1$,
$\sigma_\mathrm{Cl}+w_1$ and $\sigma_\mathrm{Na}/2+w_1$ distances, respectively.
Here, $w_1$ is the distance corresponding to the first
maximum of the distribution of solvent species, $g_{ww}(r)$ at
3.63~\AA,  approximately. Similarly, one can observe trends of formation of
clusters at a distance $(\sigma_\mathrm{Na}+\sigma_\mathrm{Cl})/2+w_2$,
where $w_2$ is the distance corresponding to
the position of the principal maximum of the distribution of solvent
species at 7~\AA,  approximately. Due to a stronger repulsion between
small cations, the formation of Na--S--Na structure is less favorable compared
to Cl--S--Cl structure at these conditions.

On the other hand, if one considers the model at $x_\mathrm{ion}=0.3$
at a point of the TMD line with slightly higher total density but substantially
lower temperature (figure~\ref{f3}) compared to the case described in figure~\ref{f2}, the
inter-ion electrostatic correlations are expected to be stronger.
Nevertheless,  screening of interactions between ions is expected to be more pronounced.
Under the conditions given in figure~\ref{f3}, one can see that the contact value
of $g_\mathrm{NaCl}(r)$ is practically of the same order as in figure~\ref{f2}.
However, the height of the shoulder of this function
at $r \approx 6.9$~{\AA} describing the correlation between oppositely
charged ions separated
by a single solvent particle is lower than in figure~\ref{f2}. On the other hand,
the height of the first maximum of $g_\mathrm{ClCl}$ is higher as an evidence
of weaker repulsion between equally charged anions.
Possible formation of Na--$w_1$--Na is also more pronounced,
cf. figure~\ref{f2}. The cation-solvent correlations are enhanced in the present
case (at this low temperature) compared
to anion-solvent correlation and compared to the case discussed in figure~\ref{f2}.

We now again return to thermodynamic manifestations of the existence of the TMD
lines. A quantitative cumulative measure of the distribution of
particles of different species is commonly given in terms of
the pair contribution to the excess internal energy per
particle as a function of the total density for different
compositions of the mixture and at different reduced temperatures.
In previous studies of the single-component soft-core model fluids
it was suggested that the relation between the density and diffusion anomalies and
the microscopic structure of the system can be explored in terms of the
excess-entropy-based framework, see e.g.,~\cite{errin}.
It has been shown that the presence of the density and the diffusion anomalies
is related to the effect of excess entropy on the fluid density.
Our discussion concerns $S_{2,\mathrm{ex}}$ defined for a mixture
by  equation (\ref{eq:s2}).

However, in contrast to previous works, we have calculated the dependence
of $S_{2,\mathrm{ex}}$ as a function of the temperature of maximum density rather than
the total density of the model. In this respect, our results given in figure~\ref{f4}~(a) correspond
to the TMD lines presented in figure~\ref{f1}. For example, the left hand part of the TMD line
for $x_\mathrm{ion}=0.1$ in figure~\ref{f1}~(b) that is characterized by $\rd T^*/\rd \rho^* > 0$ corresponds
to the upper part (lower absolute values) of the corresponding line for
$S_{2,\mathrm{ex}}$ on $T^*_\mathrm{tmd}$. What  can we learn from these functions? First
observation is that the $S_{2,\mathrm{ex}}(T^*_\mathrm{tmd})$ curves shift to lower values
of $S_{2,\mathrm{ex}}$ and lower values for $T^*_\mathrm{tmd}$ with an increasing concentration
of ionic solutes. Moreover, these curves shrink with increasing ion concentration.
At low values of $x_\mathrm{ion}$ (and at vanishing $x_\mathrm{ion}$ as well)
the curve $S_{2,\mathrm{ex}}(T^*_\mathrm{tmd})$ has a maximum such that
$\rd S_{2,\mathrm{ex}}/\rd(T^*_\mathrm{tmd}) < 0$ in the small region of $T^*_\mathrm{tmd}$ after
the point $\rd S_{2,\mathrm{ex}}/\rd(T^*_\mathrm{tmd}) = 0$. This point does not coincide
with the maximum temperature of the TMD line. Discontinuity
of this derivative is expected, however, around the point at which $T^*_\mathrm{tmd}$ reaches
its maximum. This behavior is much less pronounced at high values for ion concentration
in a mixture. The existence of such type of peculiar behavior seems to be worth to investigate
in more detail for other soft-core solvent models and for say nonpolar solutes
in various solvents for the sake of comparison with the case of ionic solutes.

A rather similar discussion can be presented while considering the dependence
of the reduced constant volume heat capacity, $C^*_V$ on $T^*_\mathrm{tmd}$, figure~\ref{f4}~(b)
[$C^*_V=C_V/(N_\mathrm{ion}+N_\mathrm{s})k_\mathrm{B}$].
The heat capacity was ``measured'' in simulations through the fluctuations of
internal energy according to equation (\ref{eq:CV}). It is quite difficult to
evaluate the heat capacity precisely due to fluctuations, especially
at low temperatures. One way to suppress the fluctuations of the $C^*_V$
values is to consider bigger boxes with a bigger number of particles, but
then the simulations become quite lengthy in time. Shift and shrinking of the
$C^*_V(T^*_\mathrm{tmd})$ curves with an increasing ion concentration are well
pronounced. However, the behavior of $C^*_V(T^*_\mathrm{tmd})$ around the extremum
point of the TMD line must be studied in more detail.

\begin{figure}[!t]
\centerline{
\includegraphics[width=0.45\textwidth,clip]{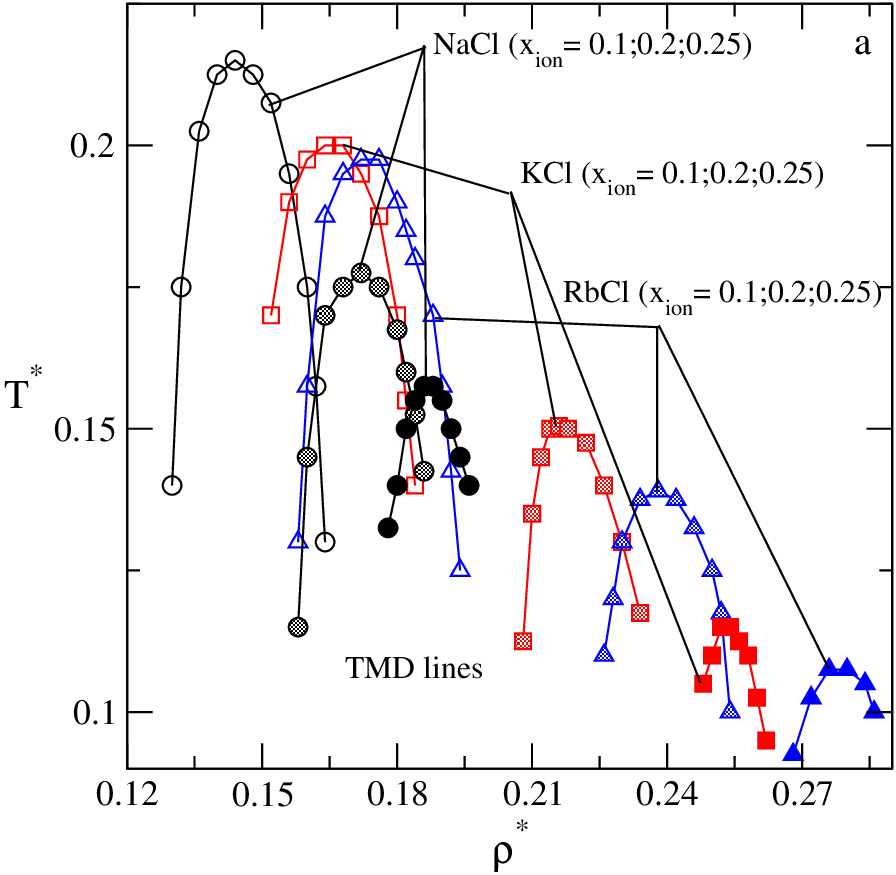}
}
\centerline{
\includegraphics[width=0.45\textwidth,clip]{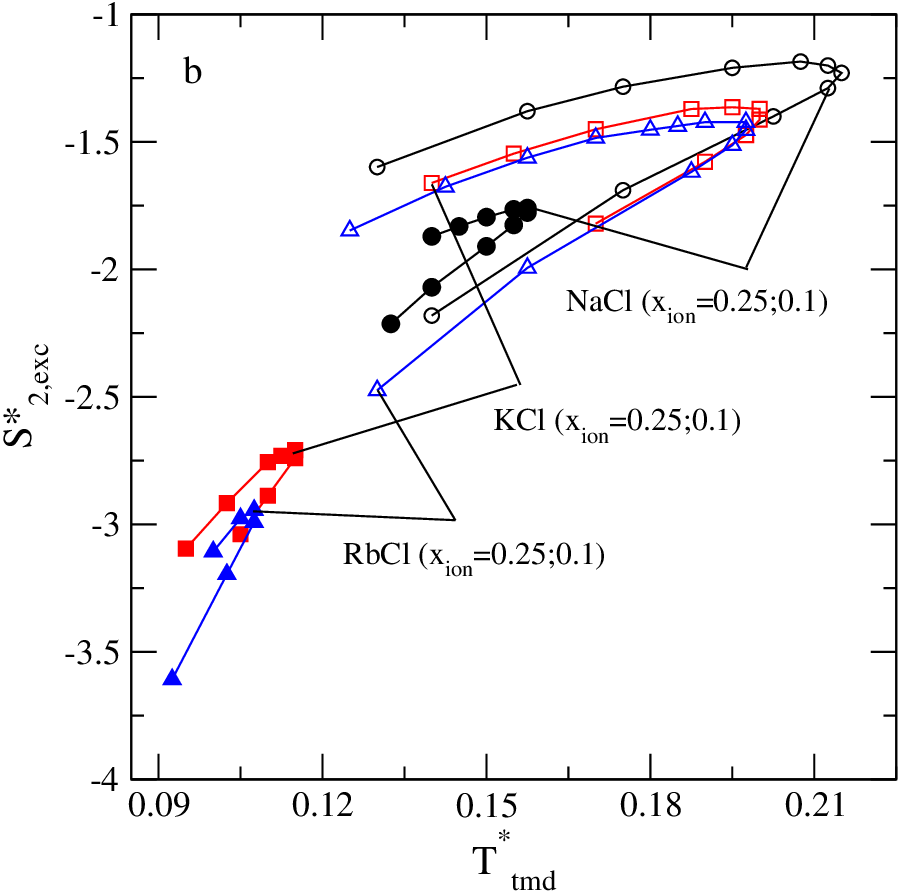}
\hspace{5mm}
\includegraphics[width=0.45\textwidth,clip]{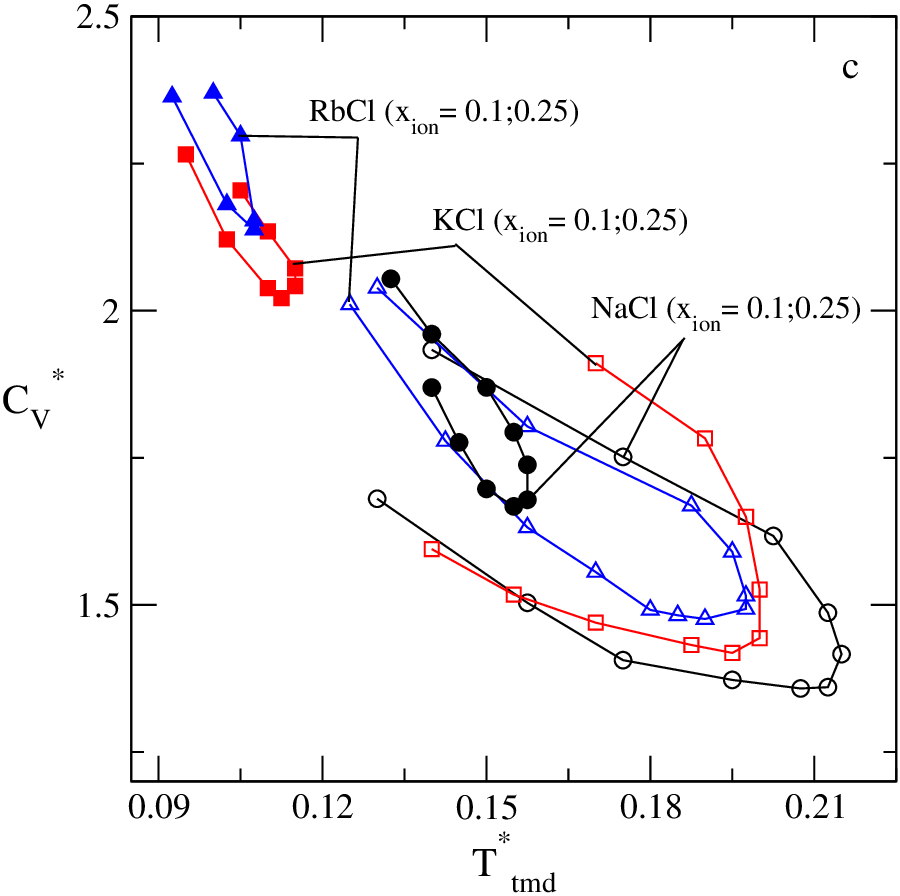}
}
\caption{(Color online) Temperature of maximum density lines of the {NaCl},
{KCl} and {RbCl} ion models with the soft-core solvent
in the $T^* - \rho^*$ plane at different concentration of ions (panel a).
The behavior of the pair contribution to excess entropy (panel b)
and of reduced heat capacity (panel c) on the
temperature of maximum density of the {NaCl},
{KCl} and {RbCl} models
at different concentration of ions.\label{f5}}
\end{figure}

After rather comprehensively examining the
properties of the TMD lines of the model NaCl salt in the soft-core solvent
in question at different ion concentrations, we would
like to proceed to the results obtained for other two salts with bigger cations,
namely the KCl and RbCl. The assumed diameters of cations were mentioned in
the previous section, but we repeat the numbers for the sake of convenience of the
reader KCl ($\sigma_+=4.74$~\AA) and RbCl ($\sigma_+=5.27$~\AA). In the former case,
the diameter of cation does not differ much from the diameter of the
anion $\sigma_\mathrm{Cl}=4.42$~\AA.  From the analysis of the $T^*-\rho^*$ projections
of the TMD lines given in figure~\ref{f5}~(a), we would like to conclude that in the present
formulation of the interactions in the model, the principal factor determining
the location of the TMD lines seems to be the excluded volume effects.
Namely, at $x_\mathrm{ion}=0.1$,
the difference between TMD lines is reasonably big, if one changes from NaCl
salt to KCl salt. The difference between KCl and RbCl cases is small
as expected, because the cation diameter does not change much.
It seems that the shift of the TMD line to higher densities is determined by the
amount of excluded volume added to the system. On the other hand, the enhanced
excluded volume effects
also lead to a lower extremal point along the temperature axis of the TMD lines.
However, this shift of TMD lines on temperature axis is to a great extent effected by
the electrostatic interactions between ions. If the ion concentration
increases for each salt model, the lines substantially shift to lower temperatures.
Unfortunately, the solvent in question (due to a set of parameters assumed) does not
exhibit vapor~-- liquid criticality. Therefore, we do not have a natural lower boundary
for the TMD lines in, for example, the $T^* - \rho^*$ projection. Our preliminary
calculations have shown that quite short TMD lines still exist for  $x_\mathrm{ion}=0.31$
and for 0.32 for NaCl model.
The shifts of the TMD lines along the density and temperature axes given in figure~\ref{f5}~(a) are
also reflected by the behavior of excess pair entropy and of the reduced heat capacity,
see figures~\ref{f5}~(b) and \ref{f5}~(c), respectively. Shrinking of these curves with an increasing
ion concentration is quite evident. Trends of the behavior of $S^*_{2,\mathrm{exc}}$ and of $C^*_V$
are similar to those discussed for the model NaCl salt in the soft-core solvent just
above.

To conclude,
in the present short communication we considered one extension of our
very recent work~\cite{miha1} concerning a simple model mixture of ions
and soft-core solvent fluid by using Monte Carlo computer simulations.
Our extension consists in considering ions of non-equal diameters and
different from the diameter of solvent particles.
We were interested to study how the presence of ion solutes affects
thermodynamic properties of the system compared to the single-component
fluid, specifically the temperature of
maximum density line. For this purpose, we studied three model solute salts, namely
the model for NaCl, KCl and RbCl.
Our results confirm that the mixtures based on a soft-core solvent in
question preserve this anomalous behavior at $x_\mathrm{ion}=0.1$ 0.2, 0.25, and 0.3,
if the cation is small, e.g., for NaCl model. However, for the model with
a big cation, e.g., RbCl the density anomaly disappears at $x_\mathrm{ion}$ around~0.27.

The TMD lines were constructed from the simulation data for a set of isochores.
They  shift to lower temperatures and higher densities with an increasing
ion concentration in the $T^* - \rho^*$ plane. The TMD line
shrinks and moves to lower values of the reduced temperature
with an increasing ion concentration for each model salt. These observations
agree with the results of molecular dynamics simulations~\cite{gallo2} for
much more elaborate model for aqueous solutions.
We have seemingly made interesting observations about the dependence
of the excess pair entropy and constant volume heat capacity along
the TMD line at different ion concentrations. Still it is necessary
to complement our study by simulations of uncharged solutes with
different characteristics mixed with applied or another soft-core model
for the solvent subsystem. It is known that in the presence of uncharged,
non-polar solutes, the curves describing the anomalies shift as well, see
e.g. \cite{pablo1}. However, for example, the TMD line changes
essentially depend on the solute concentration and strength of solute-solvent
interactions. In order to compare our present observations with the models
of uncharged solutes in soft-core solvents, additional simulations
are necessary. This project is in progress in our laboratory at present.
Intuitively, one should expect the most pronounced differences
between the effects of ions and non-polar solutes if the ion
concentration is low or, in other words, when the screening of
ion-ion interaction is not well pronounced.

There remains much room for extensions of the present work.
In order to validate the results concerning the trends for anomalies with
increasing ion concentrations, it is worth to extend the calculations and
establish the solubility limits. It is necessary to explore how the
modification of the cross solute-solvent interaction potentials would
affect the TMD lines.
Another important issue is to investigate
various properties of ionic solutions and their peculiarities
by using a solvent model characterized by the presence of liquid~-- vapor
phase transition and possibly liquid-liquid separation, e.g.,~\cite{lomba}.
Theoretical developments are in fact desirable in many of these aspects.

\section*{Acknowledgements}

O. P. has been supported in part by European Union IRSES Grant STCSCMBS 268498.
B. H.-L. acknowledges financial support of the
Slovenian Research Agency (ARRS) under grants P1-0201, and J1-4148.
O. P. is grateful to David Vazquez for technical support of this work.

\newpage

\ukrainianpart

\title
{Аномалія густини у системі заряджених твердих сфер різних розмірів
у суміші з розчинником у моделі з м'яким кором. Результати
моделювання Монте Карло}

\author{Б. Грібар-Лі\refaddr{label1}, О. Пізіо
\refaddr{label2}}
\addresses{
\addr{label1}Факультет хімії і хімічної технології, Університет
м. Любляна, 1000 Любляна, Словенія
\addr{label2} Інститут хімії,
Національний автономний університет м. Мехіко,  Мехіко, Мексика}

\makeukrtitle

\begin{abstract}
\tolerance=3000%
Недавно було досліджено вплив заряджених твердих сфер однакового
розміру як розчиненої речовини на структурні і термодинамічні
аномалії системи, в якій застосовано модель плину з потенціалом з
м'яким кором як розчинник. Застосовувався метод моделювання Монте
Карло та теорія інтегральних рівнянь  [J. Chem. Phys., 2012, \textbf{137},
244502]. Метою даної роботи є доповнити попереднє
дослідження, розглядаючи  одновалентні іони з різними діаметрами у
суміші з таким самим модельним розчинником. Зокрема, нас цікавить
аналіз  змін температури максимальної густини при збільшенні
концентрації іонів для трьох модельних солей, хлористого натрію,
хлористого калію і хлористого рубідію. З цією метою застосовуємо
моделювання Монте Карло. Наше обговорення також включає залежність
парного внеску до надлишкової ентропії та теплоємності при сталому
об'ємі вздовж лінії температури максимальної густини.  Деякі
приклади мікроскопічної структури  у вигляді парних функцій
розподілу приведено на додаток.

\keywords теорія рідин, модель плину з м'яким кором, розчин
електроліту, моделювання методом  канонічного Монте Карло, аномалія
густини

\end{abstract}

\lastpage
\end{document}